\documentstyle[12pt]{article}
\def\lapp{{\ \lower 0.6ex \hbox{$\buildrel<\over\sim$}\ }}
\def\gapp{{\ \lower 0.6ex \hbox{$\buildrel>\over\sim$}\ }}
\def\beq{\begin{equation}}
\def\eeq{\end{equation}}
\def\beqn{\begin{eqnarray}}
\def\eeqn{\end{eqnarray}}
\def\to{\rightarrow}

\def\qq{q\bar{q}}

\def\pp{{\rm p\bar{p}}}

\def\bbar{{\bar{b}}}

\def\tbar{{\bar{t}}}
\def\tt{t\bar{t}}

\def\degree{^{\circ}}

\def\GeV{{\rm GeV}}

\def\TeV{{\rm TeV}}

\def\ee{{\rm e}^+{\rm e}^-}

\def\cF{{\cal F}}

\def\iN{{1\over N}}
\def\iiN{{2\over N}}
\def\akikii{\widehat{k_1k_2}}
\def\aqiqii{\widehat{q_1q_2}}
\def\akiqii{\widehat{k_1q_2}}
\def\akiqi{\widehat{k_1q_1}}
\def\akiiqi{\widehat{k_2q_1}}
\def\akiiqii{\widehat{k_2q_2}}
\def\aqiqi{\widehat{q_1q_1}}
\def\aqiiqii{\widehat{q_2q_2}}
\def\akipi{\widehat{k_1p_1}}
\def\aqipi{\widehat{q_1p_1}}
\def\akipii{\widehat{k_1p_2}}
\def\aqipii{\widehat{q_1p_2}}
\def\akiipi{\widehat{k_2p_1}}
\def\aqiipi{\widehat{q_2p_1}}
\def\akiipii{\widehat{k_2p_2}}
\def\aqiipii{\widehat{q_2p_2}}
\def\apipii{\widehat{p_1p_2}}
\def\apiipii{\widehat{p_2p_2}}
\def\apipi{\widehat{p_1p_1}}

\def\cF{{\cal F}}

\def\antennae{1}
\def\pten{2}
\def\etafig{3}
\def\deadcone{4}
\def\deltar{5}
\def\azilg{6}
\def\alljets{7}

\begin{document}

\begin{titlepage}
\vspace*{-1cm}
\begin{flushright}
UR-1452 \\
DTP/96/08   \\
hep-ph/9605369 \\
May 1996 \\
\end{flushright}                                
\vskip 1.cm
\begin{center}                                                                  
{\Large\bf
Soft Gluon Radiation in Top Events:  Effect of Hadronic
$W$ Decays}
\vskip 1.cm
{\large Bond Masuda, Lynne H. Orr}
\vskip .2cm
{\it Department of Physics, University of Rochester \\
Rochester, NY 14627-0171, USA }\\
\vskip   .4cm
and
\vskip .4cm
{\large  W.J.~Stirling}
\vskip .2cm
{\it Departments of Physics and Mathematical Sciences, University of Durham \\
Durham DH1 3LE, England }\\
\vskip 1cm                                                                    
\end{center}                                                                    
\begin{abstract}

In reconstructing the top quark momentum from its decay products,
one must account for extra jets that can result from radiation of
gluons.  In this paper we study soft gluon radiation in top
production and decay at the Fermilab Tevatron.
We consider the cases where one or
both $W$ bosons decays hadronically, {\it i.e.}, where at least
one of the top quarks can be fully reconstructed.  We show results
for gluon
distributions at the Tevatron and find that radiation from the $W$
decay products contributes substantially in the central region.

\end{abstract}
\vfill
\end{titlepage}                                                                 
\newpage                                                                        
\section{Introduction}

The top quark's existence is now well established \cite{CDFTOP,D0TOP}
and top searches have given way to top studies.  In measuring the top quark's
properties --- in particular, its mass --- radiated gluons can play
an important role.  This is due not only to the relatively high
probability that $\tt$ events are accompanied by additional gluon jets,
but also to the fact that
gluons can be radiated in both the top production and decay
processes.  This can complicate top momentum reconstruction, leading to
ambiguities in identifying the top quark's decay products, increased
systematic uncertainties, or both.  These effects in turn
influence any measurement 
based on momentum reconstruction, and may also bias event selection based on
identifying top events via invariant mass cuts.
Therefore it is necessary to understand the distributions of radiated gluons
in top quark events.

There have been a number of studies of gluon radiation in top quark production
and decay at $\ee$ \cite{JIKIA,KOS,DKOS,SCHMIDT,OSSBIS} and 
$\pp$ colliders \cite{OSSBIS,KOSHAD,OS,LAMPE,OSS,BARGER}.
In \cite{OS} we examined additional gluons in the processes
$q \bar q, gg \to b W^+ \bar b W^-$ at the Tevatron $\pp$ collider 
in the soft gluon approximation and in
\cite{OSS} we performed the exact calculation.  These studies included 
all effects of gluons radiated from the initial state partons, from the 
$t$ and $\tbar$ quarks at the production stage, and from the $t$ and $b$
(and $\tbar$ and $\bbar$) in the $t\to bW$ decay; all effects of the 
width in the top propagator were also included.  These studies did
{\it not} include gluon radiation from hadronically decaying $W$ bosons,
so that, strictly speaking, they are directly applicable only to 
dilepton events, in which both $W$'s decay leptonically.
But in the most useful channel 
--- `lepton $+$ jets' --- only one of the $W$'s decays 
leptonically, and the other decays to quarks. 
The presence of one charged lepton in this
channel helps suppress backgrounds without sacrificing event rate, while
jets from the hadronic $W$ decay allow for direct reconstruction  of
one of the top quarks' decay products.  These events can also be accompanied by
extra gluons. In fact they are more likely to do so than dilepton events,
because there are more colored particles available to radiate.

In this paper we extend the study of Ref.~\cite{OS} for 
radiation of soft gluons  in top production and decay at the Tevatron to
include gluon radiation in hadronic $W$ decays, $W\to q\bar q'$.  In the next
section
we present analytic expressions for all contributions to the gluon radiation 
probability in the soft gluon approximation, with a discussion of properties
of the $W$ decay piece.  In section 3, we apply the results of section 2 
to $\tt$ production and decay at the Tevatron and show the resulting gluon
distributions.  Expectations for the LHC are briefly discussed.  Section 4
contains our conclusions.

\section{Soft gluon radiation in $\tt$ production and decay: general formalism}

At leading order, $\tt$ production and decay is given by the subprocess(es)
\begin{eqnarray}
a(k_1) + b(k_2) \to  t(q_1) + \bar{t}(q_2) \to
b(p_1) + W^+ + \bar{b}(p_2) + W^- \>,
\end{eqnarray}
where $ab = q\bar q$ or $gg$ for hadron colliders, and the particles' momenta 
are indicated in parentheses.  Let us assume that the $W^+$ decays
hadronically:
\begin{eqnarray}
W^+ \to q(p_3) + \bar q'(p_4).
\end{eqnarray}
If we are interested in the all jets case, then
both $W$'s decay 
hadronically and we also have 
\begin{eqnarray}
W^- \to q'(p_5) + \bar q(p_6).
\end{eqnarray}

Now consider emission of a gluon with momentum $k^\mu$.  It can come
from any of the quarks or gluons in the above processes ({\it i.e.,\/}
any of the particles whose momenta are labeled).  This includes initial
state radiation, gluons emitted by the top (or $\tbar$) quark in either the 
production or decay process,\footnote{Gluons emitted by the top quark in the 
production and decay processes do not involve separate Feynman diagrams; 
integrating over the $t$ quark's virtuality picks up poles corresponding,
respectively, to emission in the two stages.  See \cite{KOS} for a 
discussion.} and gluons emitted by the daughter quarks of the $t$ and $\bar t$
and the $W$ boson(s).

Because of the infra-red divergence associated with radiation of
gluons, an extra jet in a $\tt$ event will usually
be soft.  In the limit of soft gluon emission, several simplifications occur.
First, the gluon momentum $k^\mu$ does not affect the kinematics, so that 
all of the leading-order kinematic relations still hold.\footnote{While this
makes the computations simpler, 
it also means that results of the soft calculation
cannot be used to study the details of top mass reconstruction, which requires
an exact treatment.}
Second, the differential cross section factorizes into the lowest-order
cross section multiplied by a gluon radiation probability;
following \cite{KOS,KOSHAD,OS} we can write 
\begin{equation}
{1\over d\sigma_0}\ {d\sigma\over d E_g\> d\cos\theta_g \> d\phi_g}\ = \
{\alpha_s\over 4 \pi^2} \ E_g \ \cF \; ,
\label{softsigma}
\end{equation}
where $d\sigma_0$ is the differential cross section for the lowest-order
process ({\it i.e.}, with no gluon radiation), $E_g$ is the energy of the
soft gluon, and $\alpha_s$ is the strong  coupling.
The function $\cF$ is the sum of \lq antenna patterns' of the
radiation from the different sources listed above.
It can be written generically as
\begin{equation}
\cF = \cF_{\mbox{\tiny PROD}} + \cF_{\mbox{\tiny DEC}} +\cF_{\mbox{\tiny INT}}
\label{curlyf}
\end{equation}
where $\cF_{\mbox{\tiny PROD}}$ is the contribution from 
emission at the $\tt$ production stage (including initial-state radiation),
$\cF_{\mbox{\tiny DEC}}$ is the contribution from emission 
off the $t$ and $\bar t$
and their decay products at  the weak decay stage, 
and $\cF_{\mbox{\tiny INT}}$ is the contribution from 
interferences between these emissions. 
Because we are interested in hadronic $W$ decays, it is useful to further 
decompose $\cF_{\mbox{\tiny DEC}}$ explicitly into the contribution from
radiation off the $t$'s and $b$'s, and that from radiation off the quarks
from the $W$ decay(s):
\begin{equation}
\cF_{\mbox{\tiny DEC}}=\cF_{\mbox{\tiny DEC,tb}}+\cF_{\mbox{\tiny DEC,W}}.
\end{equation}
$\cF_{\mbox{\tiny DEC,W}}$ includes contributions from one or both $W$
decays as appropriate.

The interference term 
$\cF_{\mbox{\tiny INT}}$ contains contributions from
the interference (i) between radiation in the $t$ decay  and
that in the $\tbar$ decay and (ii) between radiation in
$\tt$ production and radiation in either decay. (As discussed below,
there is no contribution to the interference from the $W$ decays because
the $W$ boson is a color singlet.)
In principle $\cF_{\mbox{\tiny INT}}$ can have substantial effects for
gluon energies comparable to the top decay width $\Gamma_t$ 
\cite{KOS,DKOS}.
However, the observable soft jets that are relevant to the $\pp$ collider
experiments have energies much larger than $\Gamma_t$, and so in practice
the interference terms are numerically small.

Explicit expressions for  $\cF$  have been 
presented in \cite{KOSHAD,OS} for the $\qq$ and $gg$ subprocesses,
excluding hadronic $W$ decays.  The complete expressions, including 
in hadronic $W$ decays, are
\begin{eqnarray}
\cF{\mbox{\tiny PROD}}  &=& c_1 \akikii + c_2 \akiqi + 
c_3 \akiqii + c_4 \akiiqi
+ c_5 \akiiqii + c_6 \aqiqii + c_7 \aqiqi + c_8 \aqiiqii\; , \nonumber \\
\cF{\mbox{\tiny DEC,tb}}  &=& c_7 [\aqiqi + \apipi -2\aqipi ] 
+ c_8 [\aqiiqii+ \apiipii -2\aqiipii ]\; , \nonumber \\
\cF{\mbox{\tiny DEC,W}}  &=& c_9 \widehat{p_3p_4} + 
c_{10}\widehat{p_5p_6}\; , \nonumber \\
\cF{\mbox{\tiny INT}}&=& 
 \chi_1\; \left\{ c_2 [ \akipi -\akiqi ] +c_4 [ \akiipi -\akiiqi ]
 +c_6 [ \aqiipi -\aqiqii ] + 2  c_7 [ \aqipi -\aqiqi ]  \right\} \nonumber \\
 &+& \chi_2\; \left\{  c_3 [ \akipii -\akiqii ] +c_5 [ \akiipii -\akiiqii ]
 + c_6 [ \aqipii -\aqiqii ] +2  c_8 [ \aqiipii -\aqiiqii ] \right\}
  \nonumber \\
 &+& \chi_{12} \; c_6 [ \apipii -\aqipii  -\aqiipi + \aqiqii ] \; ,
\label{general}
\end{eqnarray}
where the functions $\widehat{pq}$  and $\chi_i$ 
are defined by
\begin{eqnarray}
\widehat{pq} \> & = & \> {p\cdot q \over p\cdot k \ q\cdot k} \; ,\nonumber \\
\chi_i & = & {m_t^2\Gamma_t^2 \over (q_i\cdot k)^2 + m_t^2 \Gamma_t^2 } 
\qquad (i=1,2)\; , \nonumber \\
\chi_{12} & = & {m_t^2\Gamma_t^2\; (q_1\cdot k\; q_2\cdot k + m_t^2\Gamma_t^2)
 \over \left[ (q_1\cdot k)^2 + m_t^2 \Gamma_t^2 \right]
    \; \left[ (q_2\cdot k)^2 + m_t^2 \Gamma_t^2 \right] } \; .
\label{chidef}
\end{eqnarray}
In terms of the gluon energy, we have $\cF \sim E_g^{-2}$, and so
the cross section (Eq.~(\ref{softsigma})) has the infra-red 
behavior  $d \sigma / d E_g \sim E_g^{-1}$, as expected. 
Additional collinear
singularities arise from the $p\cdot k$ denominators when $p^2 = 0$.

The coefficients $c_i$ depend on the color structure of the
hard scattering and can be different for the $\qq$ and $gg$ processes
($C_F = 4/3,\ N=3$):
\begin{center}
\begin{tabular}{|c|c|c|}  \hline                                    
\rule[-1.2ex]{0mm}{4ex}  coefficient & $\qq\to\tt$ & $gg\to\tt$ \\ \hline 
 $c_1$       &  $-\iN$   &   $-2C_F+2N+2Y$    \\
 $c_2$       &  $2C_F-\iN$   & $C_F-X-Y$     \\
 $c_3$       & $\iiN$    &  $C_F+X-Y$    \\
 $c_4$       & $\iiN$    &  $C_F+X-Y$    \\
 $c_5$       & $2C_F-\iN$    &    $C_F-X-Y$    \\
 $c_6$       & $-\iN$    &  2Y     \\ \hline
 $c_7, c_8$       & $-C_F$    &  $-C_F$    \\
 $c_9, c_{10}$       & $2C_F$    &  $2C_F$    \\
  \hline                                                                        
\end{tabular}
\end{center}
Note that the coefficients $c_7$--$c_{10}$ are involved only in the decay
contributions (including decay--decay interference) and are therefore 
independent of the production process.
The quantities $X$ and $Y$ depend on the $gg\to\tt$ subprocess energy
and scattering angle. Explicit expressions and a discussion can be found in
\cite{KOSHAD}.

The production, $tb$ decay, and interference contributions to $\cF$ 
have been studied
in Refs.~\cite{KOSHAD,OS}, to which the reader is referred for further
discussion.  
We shall focus here on the $W$ decay contribution.
Consider a single hadronically decaying $W$:  $W\to q \bar q'$.
Its most important feature, for our purposes, is that the $W$ boson is a 
color singlet.  Radiation from its decay products at $O(\alpha_s)$
             does not, therefore,
interfere with gluons from other sources in the process.  Its contribution 
to the radiation pattern simply adds incoherently to the total at this order.

In fact this contribution is just the familiar quark-antiquark antenna
pattern, 
whose properties are well known and are discussed, {\it e.g.,\/} in 
Ref.~\cite{BOOK}.  Its color structure is the same as for the
decay $Z\to q\bar q$.
Here we remind the reader of a few of its properties.
Because the $q$ and $\bar q$ from the $W$ decay are taken to be massless,
their
contribution to $\cF$ depends only on their orientation relative to each
other and to the gluon.  Furthermore, the only other kinematical factor is an 
overall
$E_g^{-2}$, which is common to all terms in $\cF$.  Explicitly,
\begin{eqnarray}
\cF{\mbox{\tiny DEC,W}}  &=& 2 C_F {p_3\cdot p_4 \over p_3\cdot k \ p_4\cdot k}
\; \nonumber \\
&=& 2 C_F {(1-\cos\theta_{34}) \over E_g^2 (1-\cos\theta_3)(1-\cos\theta_4)}
\; ,
\label{antenna}
\end{eqnarray}
where $\theta_3$, $\theta_4$, and $\theta_{34}$ are, respectively, the angles
between the gluon and the quark, the gluon and the antiquark, and the quark
and the antiquark.  Figure~{\antennae} shows the distribution given in
Eq.~(\ref{antenna}) as a function of gluon angle for several 
$q\bar q$ angles.  We see the collinear singularities characteristic of
radiation from massless particles.  In fact most (but not all) of the radiation 
is concentrated in the immediate vicinity of the two quarks, which translates 
in practice to strong sensitivity to jet separation cuts.  We also 
see the string effect --- the 
enhancement of radiation in the region between the $q$ and $\bar q$ 
compared to that outside them.  
This enhancement can be quite significant when the 
quarks are relatively close together in angle.  As the angle increases, the 
enhancement decreases, but does not disappear altogether, and for
quarks that are back-to-back (Fig.~1(d)) we see the plateau characteristic of
initial state radiation, where the radiating particles are also back-to-back.

Finally we note that the preceding discussion of the $W$ decay contribution to
the gluon radiation is true {\it independent} of the source of the 
$W$, as a consequence of its being a color singlet object.
In particular, although we are concerned here with hadronic top production
at the Fermilab Tevatron collider, hadronic $W$ decays in top production
at, for example, an $\ee$ collider, are taken into account in exactly 
the same way.
At the order to which we are working,
the $W$ decay antenna simply adds incoherently to the contribution from the
remaining sources of gluon radiation in any process of
interest.\footnote{Perturbative interference between radiation off the
$W$ and other colored particles in the scattering process begins at
$O(\alpha_s^2)$ with the exchange of two gluons in a color singlet
state. This has been studied in the context of $e^+e^- \to W^+W^-
\to q \bar q q \bar q$ production \cite{MWIGPZ,MWISK}.
In general, such interference is suppressed by factors of
$1/N_c^2$ and $(\Gamma_W/\omega)^2$, where $\omega$
is a scale characteristic of the gluon energies,
 and is therefore numerically very small
\cite{MWISK}.}

\section{Gluon distributions in $\tt$ events at the Tevatron}

\subsection{Single hadronic $W$ decay:  lepton $+$ jets mode}

The soft gluon radiation pattern described by Eq.~(\ref{softsigma})
can be combined with the lowest-order cross section to obtain the
full gluon distribution in a particular experimental situation.  In 
this section we present the resulting gluon distributions for $\tt$ 
production at the Fermilab Tevatron $\pp$ collider for the 
case where one $W$ decays leptonically and the other decays hadronically,
{\it i.e.,} the `lepton $+$ jets' mode.
As we have seen in the previous section, the contribution from the 
hadronic $W$ decay adds incoherently to the other contributions.
The latter correspond to what we would have for the dilepton mode, {\it i.e.,}
to the results in Ref.~\cite{OS}.
We note that results for the production and `$tb$' decay distributions 
presented 
below are virtually the same as in \cite{OS}, where a more complete
discussion of them can be found.  Our emphasis here will be on the 
new features resulting from the $W$ decay contribution.

We consider the process $p\bar p\to\tt\to b l \nu \bar b q \bar q $, with
gluons generated according
to Eq.~(\ref{softsigma}), with $c_{10}=0$ in Eq.~(\ref{general}),
for $\pp$ center-of-mass energy 1.8 TeV.
We take $m_t=174\ {\rm GeV/c^2}$, $M_W=80\ {\rm GeV/c^2}$, and 
$m_b=5\ {\rm GeV/c^2}$
and we use MRS(A$'$) parton distributions \cite{MRS}.
We work at the parton level, with kinematic cuts corresponding roughly
to those in the experiments:  central production of leptons
and quarks, some minimum separation of jets from each other and 
from the lepton, and energy and transverse momentum cuts appropriate to 
detectability of jets but also consistent with the soft gluon approximation.
The cuts are 
\begin{eqnarray}
|\eta_{i}| \> & \leq & \> 1.5 \; ,\nonumber \\
|\eta_g| \> & \leq & \> 3.5 \; ,\nonumber \\
10\ \GeV/c \leq \> & p_T^g & \> \leq  25\ \GeV/c \; ,\nonumber \\
E_g \> & \leq & \> 50\ \GeV \; ,\nonumber \\
10\ \GeV/c \leq \> & p_T^i  \; ,\nonumber \\
\Delta R_{ig},\Delta R_{ij} \> & \geq & \> 0.5 \; ,
\label{cuts}
\end{eqnarray}
where $\eta$ is the pseudorapidity and $(\Delta R)^2 =
(\Delta\eta)^2 + (\Delta\phi)^2$. In Eq.~(\ref{cuts})
$g$ represents the gluon jet and $i$ and $j$
represent any other detected jet ($b$, $\bar b$, or either of the 
quarks from the $W$ decay)  or the charged lepton.
The $\eta_g$ and $\Delta R_{ig}$
cuts eliminate the collinear singularities associated with emission from 
massless particles.

The resulting distributions in the gluon energy and 
transverse momentum are shown as solid histograms in Fig.\ {\pten}.  
The contribution associated
with top production is shown as a dotted line, that for 
$\cF_{\mbox{\tiny DEC,tb}}$ as a dashed line, and that for 
$\cF_{\mbox{\tiny DEC,W}}$ as a dot-dashed line.  (Interference terms are
included in the total but are too small to appear on the plots.) From 
the figure we
see two features of the $W$ decay contribution that will also appear
in the rapidity distribution, {\it viz.,\/} that it is comparable in both size
and shape to the remaining decay piece.  
Recall that the latter comes from
the decays of both the $t$ and the $\bar t$, whereas the former
involves only the decay of a single top.

The gluon pseudorapidity distribution is  shown in Fig.~{\etafig};
again the $W$ decay piece looks very similar
to the `$tb$' piece.  We conclude that hadronic $W$ decays contribute
substantially to gluon radiation in the central 
rapidity region, more than doubling the amount of radiation associated
with the top quark decays.

For purposes of distinguishing the production and decay contributions 
in the dilepton case, we also considered in \cite{OS} the behavior of the gluon 
distribution as a function of angular distance $\Delta R_{bg}$ of the
gluon  from the 
$b$ quark.  In Fig.~{\deadcone}(a) we show the distribution in
$\Delta R_{bg}$ for very small angles ($\Delta R<0.5$).  Except for allowing
$\Delta R_{bg} \geq 0.01$, 
we have retained the cuts specified in Eq.~(\ref{cuts}).
We see that in
this region the distribution is dominated by the `$tb$' decay contribution
(dashed histogram) and that as $\Delta R$ decreases, the distribution
increases and then turns over, displaying the `dead-cone' behavior 
characteristic of radiation from massive particles \cite{BOOK}.  In contrast,
the distribution in the angular distance $\Delta R_{Wjet,g}$ 
between the gluon and one of the (massless) quarks from the hadronic $W$ decay,
shown in Fig.~{\deadcone}(b), should have a singularity when the quark and
gluon are collinear.  
We see that when we reduce the minimum allowed $\Delta R_{Wjet,g}$ to $0.01$,
the distribution 
shows no signs of turning over at small angles.
The singularity is avoided only by the $\Delta R$ cut.

While it is true that Fig.~{\deadcone} shows the difference between 
emissions from massless and massive particles, the effect occurs at 
angles much too small for the difference to be detected experimentally.
A less academic view appears in Fig.~{\deltar}(a) and (b), where
we look at the same distributions on a larger scale after 
reimposing the $\Delta R > 0.5$ requirement.  The distributions look remarkably 
similar, the only real difference being which contribution
dominates at small angles:  the `$tb$' contribution (dashed histogram) in
Fig.~{\deltar}(a) and the $W$ decay contribution (dot-dashed histogram)
in Fig.~{\deltar}(b).
This could be used in principle to distinguish 
between the two decay contributions if one knew, from secondary vertex
detection, for example, which were the $b$ jets and which were those from 
the $W$ decays (assuming the gluon jet to be the softest one).
However the results are quite sensitive to both the $\Delta R$ cuts
and to fragmentation, which is not included in our parton-level analysis.

A perhaps more promising way to separate out the $W$ decay contribution 
is to note that it is associated with only {\it one} of the top decays.
If the top quarks are produced more or less back-to-back,  then
some of that separation can be expected to 
survive the decay process.  In that case the charged  
lepton would tend not to be too close to the quarks from the other $W$ decay.
And since gluons prefer to be near the quarks that radiated them, 
radiation from the hadronic $W$ decay should increase 
with angular distance from the charged lepton.  
The expected effect is quantified in Fig.~{\azilg},
where we show the gluon distribution as a function of the azimuthal
angle difference between the lepton and the gluon.  We consider the azimuthal
angles because the $t$ and $\bar t$ are produced back-to-back in the 
transverse plane.  In addition to the cuts of Eq.~(\ref{cuts}), we require
the angular separation of the $b$ and $\bar b$ to be greater than $45\degree$,
and that all detected particles (lepton
and all jets, including the soft gluon jet)
have $|\eta|<1$.  These additional cuts are designed to increase
the likelihood that the event is more central and the $t$ decay products 
more widely separated from those of the $\bar t$.  The pseudorapidity cut 
on the gluon has the added advantage that it eliminates a large 
portion of the contribution from production ({\it cf.} Fig.~{\etafig}).
In Fig.~{\azilg} we see the expected behavior.  
All of the distributions are suppressed below about $30\degree$ 
because of the $\Delta R$ cut.  Beyond that, 
the production and
`$tb$' decay contributions are relatively flat.
In contrast, the $W$ decay 
contribution (dot-dashed histogram) starts out at about the same size 
as the other two and increases with increasing angle, reaching a maximum
about 50\% higher than the initial value.  {\it All} of the increase in the 
total gluon distribution is due to the increase in the $W$ decay
contribution.

Similar effects would be expected in the distribution in 
azimuthal
angle difference between the gluon and one of the $b$ quarks, due to the
typical angular separation of about $\pi/2$ between the $b$ and $W$
from a single top decay.  However, using the lepton is more practical
experimentally because there is only one lepton 
in each event,\footnote{Not counting those from semileptonic $b$ decays, which 
can presumably be eliminated with an isolation cut.}
and also because identifying it and measuring its momentum
does not involve the complications associated with doing so for
jets.

\subsection{Two hadronic $W$ decays:  all jets mode}

If both the $W^+$ and $W^-$ in a $t \bar t$ event decay hadronically, then
there are six jets 
in the final state at leading order.  The advantages of this mode are 
its large branching ratio and the fact that, in principle, both the $t$ and 
$\bar t$ can be fully reconstructed from their decay jets.  In practice,
combinatorics and huge QCD backgrounds make this task more difficult.  
In addition, radiation of a gluon can give rise to an extra
jet, further complicating matters.  

If the additional gluon is soft, 
its distribution is given by Eq.~(\ref{softsigma}) with the coefficient 
$c_{10}=2C_F$ in $\cF_{\mbox{\tiny DEC,W}}$ as given in the table.  
Now, in addition to the $W^+$ decay contribution discussed above, we 
also have one from the hadronic decay of the $W^-$.  
This new contribution is the same as that from the $W^+$, and it
adds incoherently.  Therefore 
radiation from the $W$ decays is effectively doubled when both $W$'s decay
to quarks.  This is illustrated in Fig.~{\alljets}, which shows the 
gluon rapidity distribution at the Tevatron, where both $W$'s contribute
to the $W$ decay piece (shown as a dot-dashed histogram).  The cuts are
as in Eq.~(\ref{cuts}).  Their effect here is slightly different from above,
however, because now there is no            charged lepton, and the
lepton and neutrino
(to which no cuts were applied previously) are replaced by two quarks.
The jet cuts now apply to {\it all} of the particles in the final state.
This accounts for the fact
that the $W$ decay contribution in the all jets case is slightly less than
twice that for lepton $+$ jets.  The transverse momentum, energy, and 
$\Delta R$ distributions exhibit a similar doubling of the $W$ decay 
contribution, and are omitted for brevity.

\subsection{Expectations for the LHC}

The CERN Large Hadron Collider (LHC) will produce large numbers of top quark
pairs.
Gluon radiation in top production and decay at the LHC 
differs from that at the Tevatron largely because of the difference in 
collision energies.  At the higher-energy LHC, the top quarks are more
energetic, and 
the cross section is dominated by the gluon-gluon initial state, in contrast
to the quark-antiquark annihilation that dominates at the Tevatron.
The result \cite{OSSLHC} is a much larger proportion of production-stage
radiation at the LHC than the Tevatron, with relatively little (on the 
order of 10--20\%) radiation occurring in the decays in the dilepton case.
We expect the effects
of radiation from hadronic $W$ decays to be similar to effects at the Tevatron:
radiation from a single hadronically decaying $W$ would
be comparable to the total decay contributions from the $tb$ and 
$\bar t \bar b$ antennae.  Production-stage radiation would continue to 
dominate. We defer a detailed discussion to an exact treatment.

\section{Conclusions}

We have presented results for soft gluon radiation in top quark production and
decay at the Fermilab Tevatron for the cases relevant for direct reconstruction
of the top quarks:  the lepton $+$ jets and all jets decay modes.  
Allowing for radiation at $O(\alpha_s)$ from hadronic $W$ decays
is straightforward because
the $W$ is a color singlet.  The corresponding contribution simply
adds incoherently to the total, with no interference with the remaining
contributions.  For this reason, soft radiation in $W$ decays is also easily 
treated in top production in $e^+e^-$ collisions by adding 
$\cF_{\mbox{\tiny DEC,W}}$ to the $\cF$ appropriate to the electron-positron 
initial
state ({\it cf.} Eqs.~(\ref{softsigma},\ref{curlyf})).

We found that, at the Tevatron, gluons radiated in hadronic 
$W$ decays contribute subtantially to the total amount of radiation
in the central region.  In particular, gluon radiation from {\it each} 
hadronic $W$ decay is comparable both in size and distribution to 
the {\it total} $tb$ and $\bar t \bar b$ decay contributions.  
The $W$ and non-$W$ contributions are thus difficult to distinguish.
But by considering the distribution in the azimuthal angle between the gluon
and the charged lepton in the lepton $+$ jets mode
we saw that some separation is possible.
We also note that since the $tb$ and $\bar t \bar b$ decay contributions
comprise the total decay contribution in the dilepton
detection mode, the $W$ decay contribution could in 
principle be isolated by making a direct comparison of jet distributions in 
dilepton and lepton $+$ jets events.  

In attempts to measure the top mass by reconstructing the momenta of its 
decay products, it is very important to take into account the presence of 
extra jets due to gluon radiation.  As we have shown in the soft 
approximation, hadronic $W$ decays 
give rise to significant amounts of gluon radiation.  An analysis of the 
impact of this radiation on top momentum reconstruction requires an 
exact treatment, which we plan in future work.

\bigskip
\medskip                   
\noindent{\Large\bf Acknowledgements}
\bigskip

\noindent   
This work was supported in part by the U.S.\ Department of Energy,
under grant DE-FG02-91ER40685, and by the University of Rochester SummerReach
Program.
\goodbreak

\vskip 1truecm
\bibliographystyle{unsrt}

\vskip 1truecm

\section*{Figure Captions}
\begin{itemize}
\item [{[\antennae]}]
Single $W$ decay contribution $\cF_{\mbox{\tiny DEC}}$ (Eq.~(\ref{antenna}))
to the soft gluon radiation pattern in the $\theta - \phi$ plane, in
arbitrary units, for  $\theta_{34}=$ (a) $45\degree$, (b) $90\degree$,
(c) $135\degree$, and  (d) $180\degree$.

\item [{[\pten]}]
Distributions in (a) the gluon energy and (b) the gluon transverse momentum,
in $\tt$ production, via the subprocesses
$q \bar q, gg \to b W^+ \bar b W^-$, in $\pp$ collisions at $\sqrt{s} =
1.8\ \TeV$, with $W^+\to q \bar q$.  
Contributions from production (dotted lines), 
$\cF_{\mbox{\tiny DEC,tb}}$ decay (dashed lines), and
$\cF_{\mbox{\tiny DEC,W}}$ (dot-dashed lines)
are shown along with their totals (solid lines).  The cuts are
listed in  Eq.~(\ref{cuts}).

\item [{[\etafig]}]
Gluon pseudorapidity distributions in $\tt$ production, via the subprocesses
$q \bar q, gg \to b W^+ \bar b W^-$, in $\pp$ collisions at $\sqrt{s} =
1.8\ \TeV$, with $W^+\to q \bar q$.  
The net distribution is shown as a solid line; contributions 
from production (dotted lines),
$\cF_{\mbox{\tiny DEC,tb}}$ decay (dashed lines), and
$\cF_{\mbox{\tiny DEC,W}}$ (dot-dashed lines)
are also shown.  The cuts are
listed in  Eq.~(\ref{cuts}).

\item [{[\deadcone]}]
Distribution in $\Delta R$ between the gluon and (a) the $b$ quark and 
(b) the charged lepton.  Cuts are as in Eq.~(\ref{cuts}), except for
allowing (a) $\Delta R_{bg} \geq 0.01$ and (b) $\Delta R_{W jet,g} \geq 0.01$.

\item [{[\deltar]}]
Distribution in $\Delta R$ between the gluon and (a) the $b$ quark and
(b) the charged lepton.  The cuts are as in Eq.~(\ref{cuts}).

\item [{[\azilg]}]
Distribution in the azimuthal angle between the charged lepton and gluon jet.
The cuts are as in Eq.~(\ref{cuts}), except $|\eta|<1$ for all detected
particles and the angle between the $b$ and $\bar b$ must be greater
than $45\degree$.

\item [{[\alljets]}]
Gluon pseudorapidity distributions in $\tt$ production, via the subprocesses
$q \bar q, gg \to b W^+ \bar b W^-$, in $\pp$ collisions at $\sqrt{s} =
1.8\ \TeV$, with both $W$'s decaying hadronically.
The net distribution is shown as a solid line; contributions 
from production (dotted lines),
$\cF_{\mbox{\tiny DEC,tb}}$decay (dashed lines), and
$\cF_{\mbox{\tiny DEC,W}}$ (dot-dashed lines)
are also shown.  The cuts are
listed in  Eq.~(\ref{cuts}).

\end{itemize}

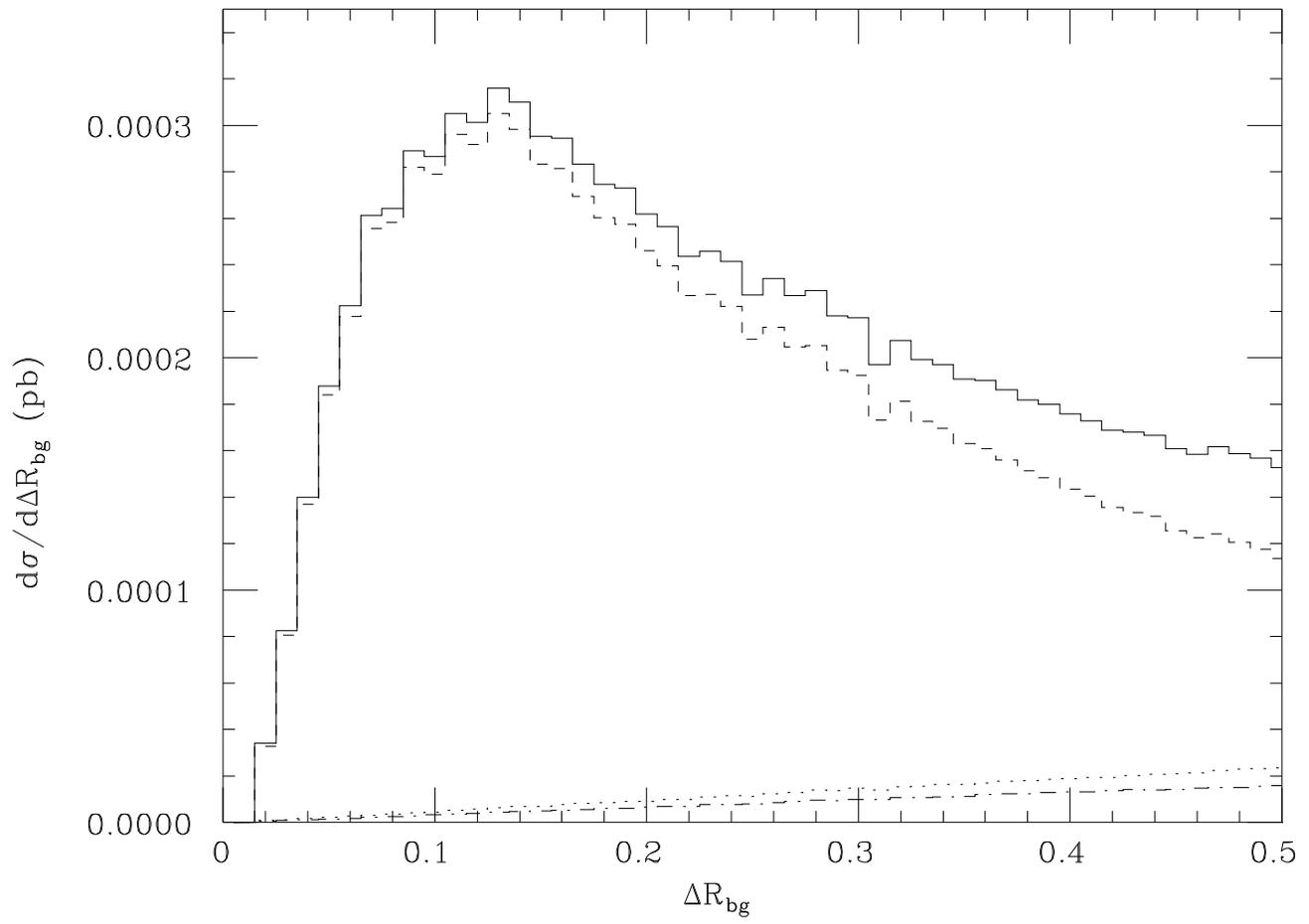
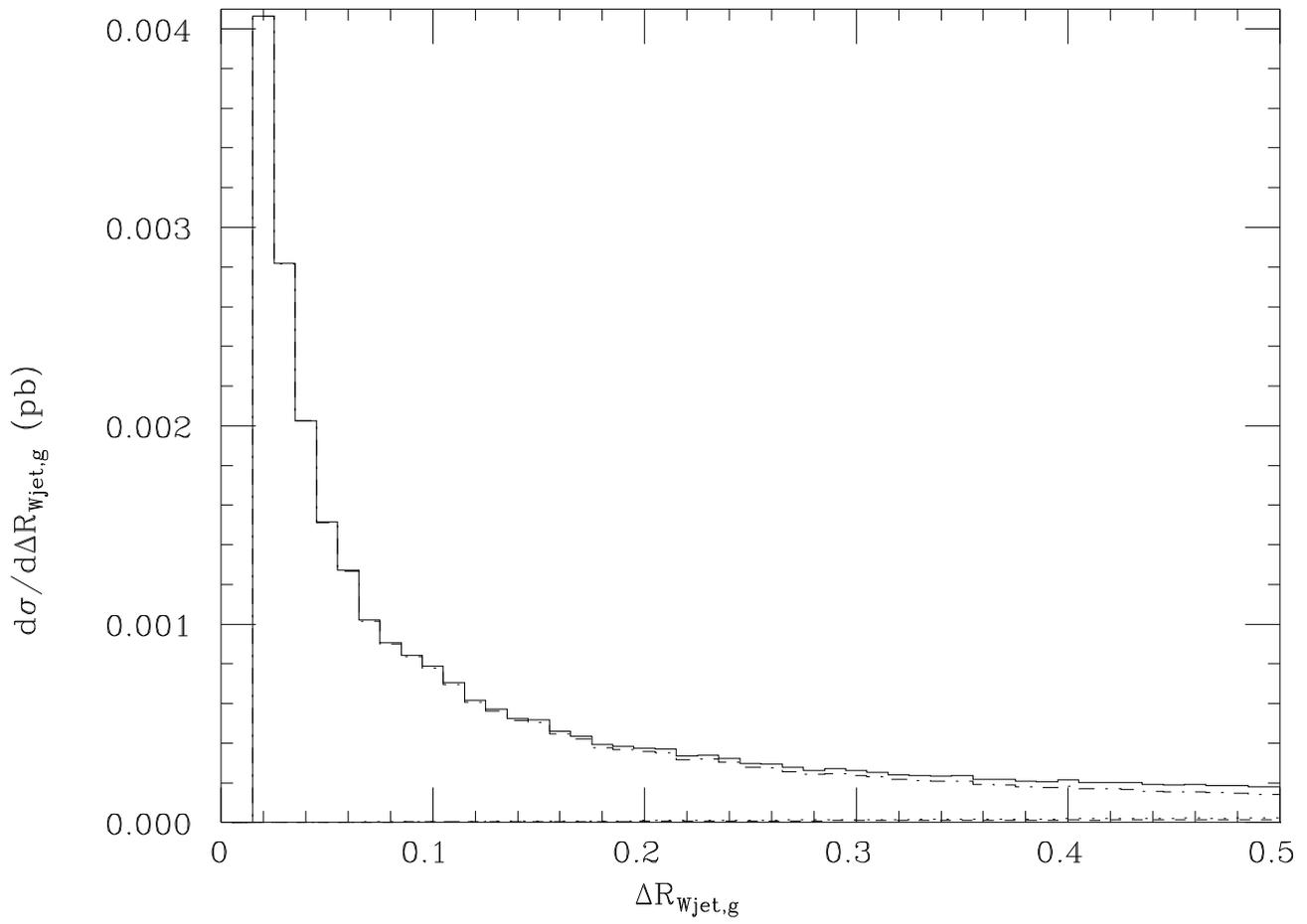
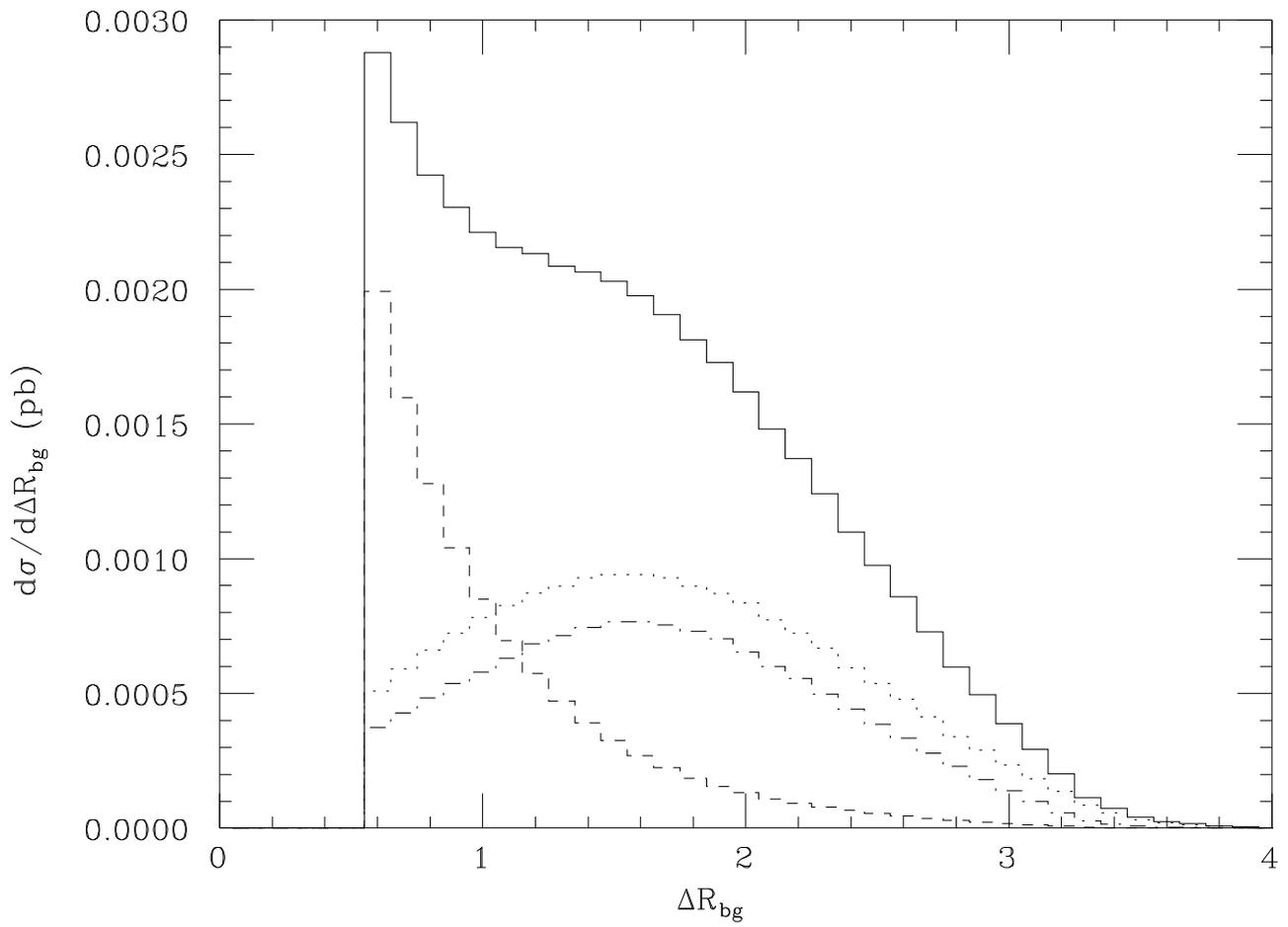
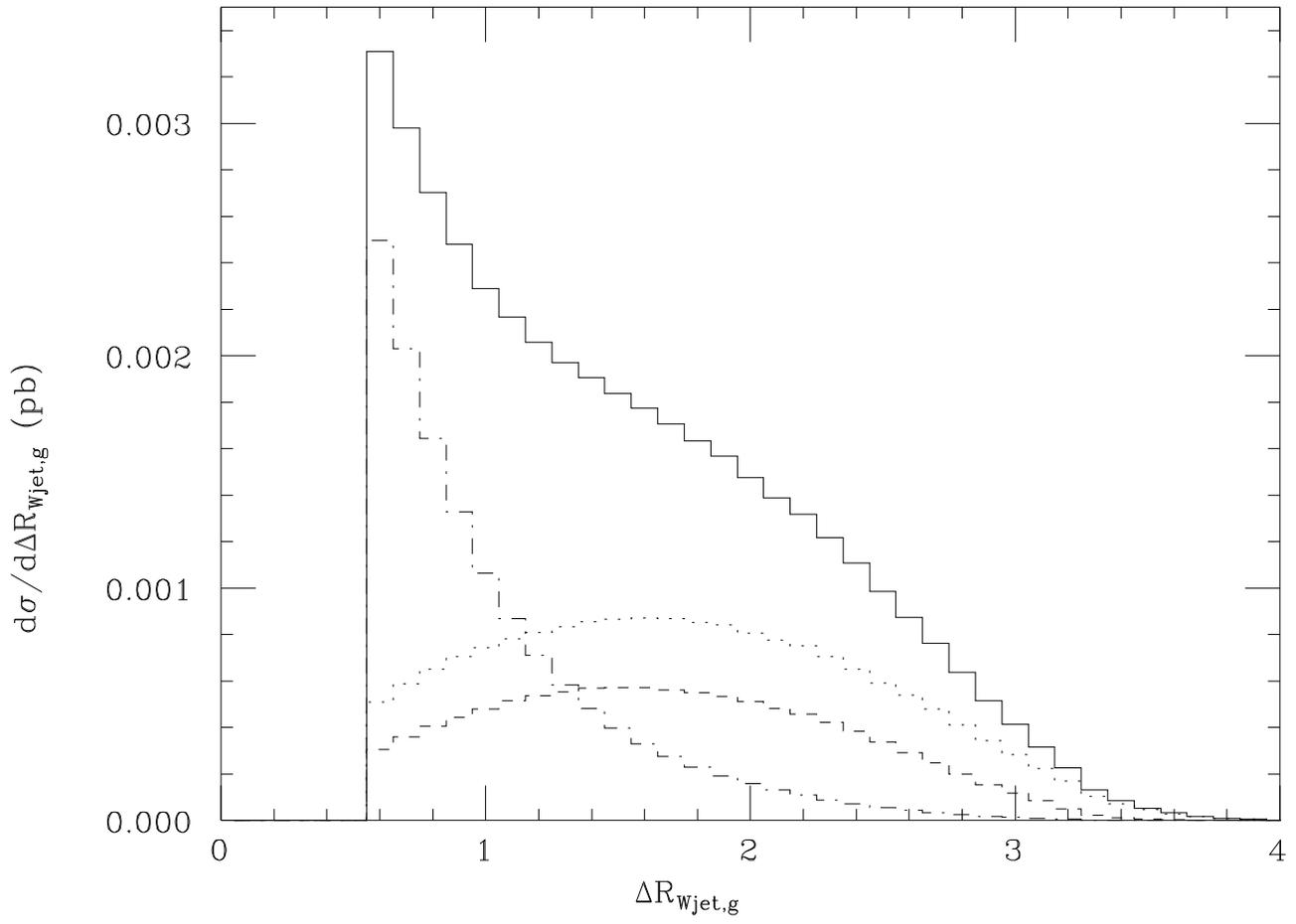
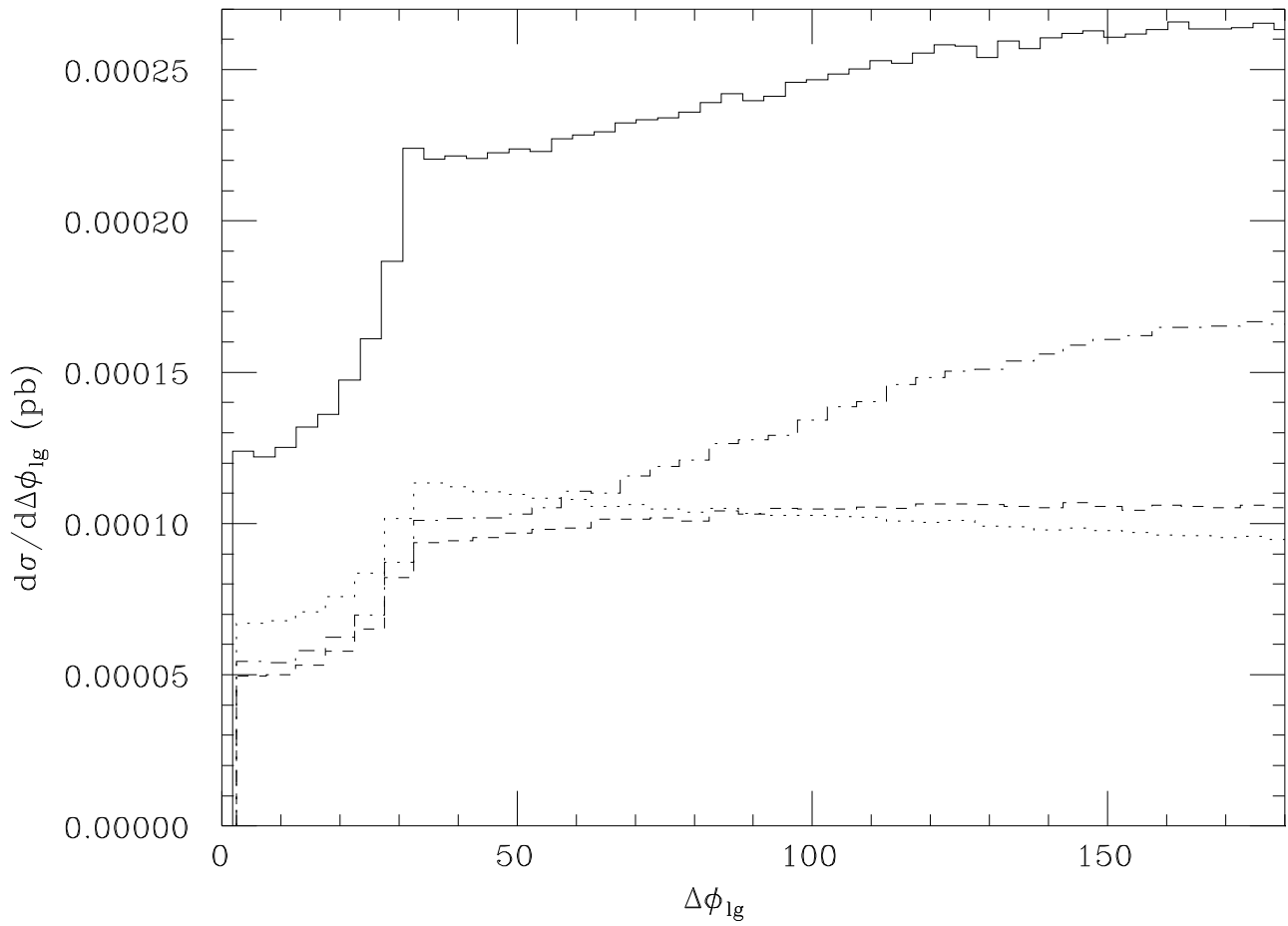
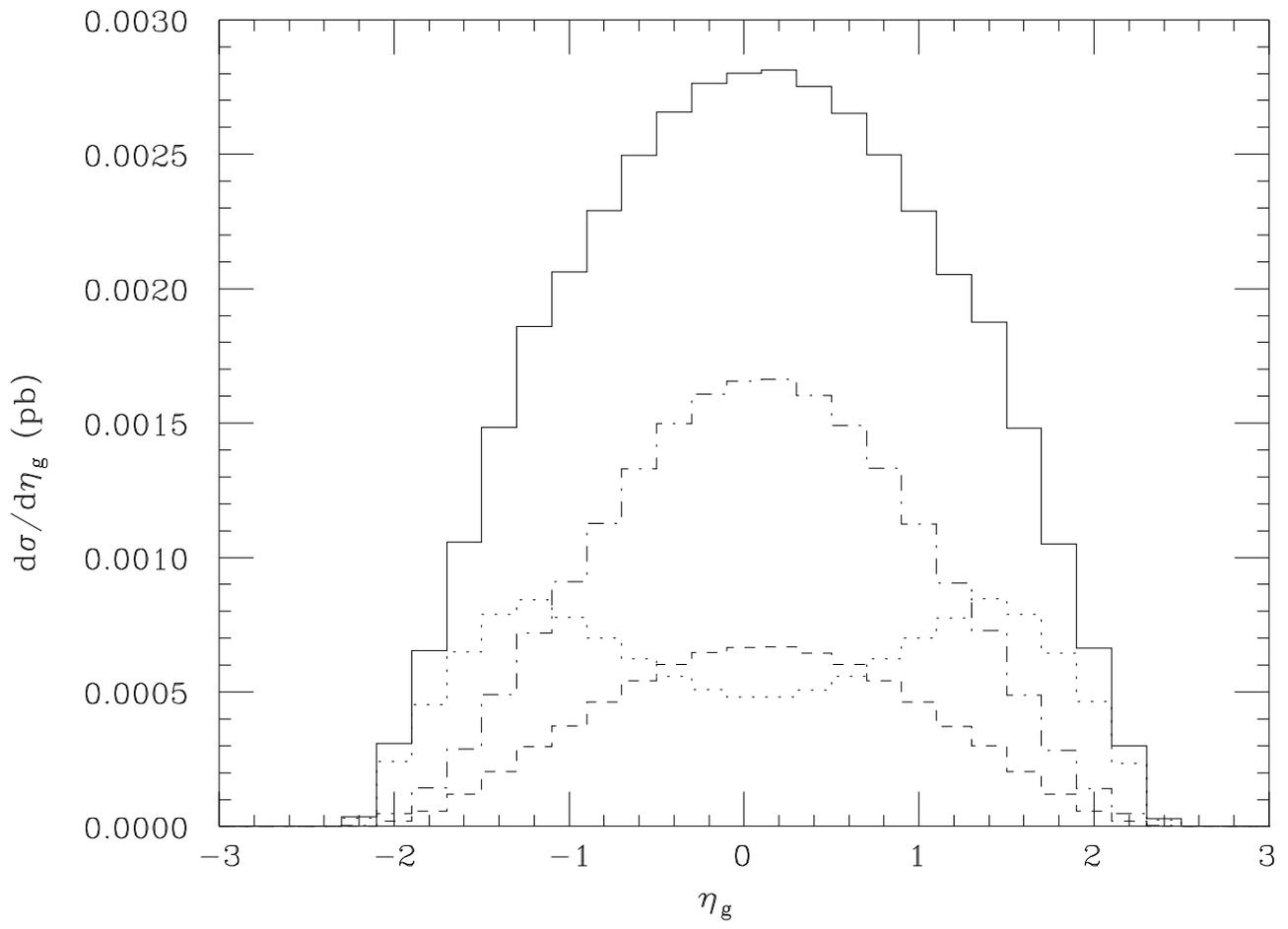
\end{document}